\documentclass[11pt]{article}

\usepackage{amssymb,amsmath,amsfonts,amsthm,lscape}

\usepackage{graphicx}

\newcommand{\bq}{\mathbf{q}}
\newcommand{\bp}{\mathbf{p}}

\newcommand{\btq}{ \mathbf{q}}

\newcommand{\tq}{ {q}}

 \newcommand{\kk}{\kappa}
   
  \newcommand{\Om}{\Omega}

 \def\k{{\kappa}}

\def\1{\'{\i}}                           

  \def\>#1{{\mathbf#1}}

   \def\m{\mu}

\begin{document}

\thispagestyle{empty}

\

 \vskip1cm

\noindent {\Large{\bf {A curved H\'enon--Heiles system and its integrable perturbations
}}}

\medskip
\medskip
\medskip
\medskip

\begin{center}
{\sc \'Angel Ballesteros, Alfonso Blasco\footnote{
 Based on the contribution presented at ``The 30th International Colloquium on Group Theoretical Methods in Physics",
July 14--18, 2014, 
 Ghent, Belgium. To appear in {\em Journal of Physics: Conference Series}.} and Francisco J. Herranz
}

\medskip
\medskip

\noindent
{Departamento de F{\'{\i}}sica, Universidad de Burgos, E-09001 Burgos, Spain}

\medskip

\noindent{E-mail:\quad {\tt angelb@ubu.es, ablasco@ubu.es, fjherranz@ubu.es}}

  \end{center}

\medskip

\medskip

\begin{abstract}
\noindent
The constant curvature analogue   on the two-dimensional sphere  and  the hyperbolic  space of the integrable H\'enon--Heiles   Hamiltonian $\mathcal{H}$ 
   given by
$$
\mathcal{H}=\dfrac{1}{2}(p_{1}^{2}+p_{2}^{2})+ \Omega \left(  q_{1}^{2}+ 4 q_{2}^{2}\right) +\alpha \left(
q_{1}^{2}q_{2}+2 q_{2}^{3}\right) ,
$$
where  $\Omega$ and  $\alpha$   are real constants, is revisited. The resulting integrable curved Hamiltonian, $\mathcal{H}_\k$,   depends on a   parameter $\k$ which is just the   curvature of the  underlying space and allows one to recover 
 $\mathcal{H}$  under the smooth  flat/Euclidean limit $\k\to 0$. This system can be regarded as an integrable cubic perturbation of a specific curved $1:2$ anisotropic oscillator, which was already known in the literature.  The Ramani-Dorizzi-Grammaticos  (RDG) series of potentials associated to $\mathcal{H}_\k$ is fully constructed, and corresponds to the curved integrable analogues of homogeneous polynomial perturbations of $\mathcal{H}$ that are separable in parabolic coordinates. Integrable perturbations of   $\mathcal{H}_\k$ are also fully presented,  and they can be regarded as the curved counterpart of  integrable {rational} perturbations of 
the Euclidean Hamiltonian  $\mathcal{H}$. It will be explicitly shown that the latter perturbations can be understood as  the `negative index' counterpart of the curved RDG series of potentials. Furthermore, it is shown that  the integrability of the curved H\'enon--Heiles Hamiltonian $\mathcal{H}_\k$ is preserved under the simultaneous addition of curved analogues of `positive' and `negative' families of RDG potentials.

\end{abstract}

\newpage

%%%%%%%%%%%%%%%%%%%%%%%%%%%%%%%%%%%%%%%%%%%%%%%%%%%%

\section{Introduction}

The H\'enon--Heiles Hamiltonian 
$$
{H}=\dfrac{1}{2}(p_{1}^{2}+p_{2}^{2}) + \dfrac{1}{2}(q_{1}^{2}+q_{2}^{2})+\lambda\left(
q_{1}^{2}q_{2}-\frac{1}{3}\,q_{2}^{3}\right)
\label{HHaut}
$$
was introduced in~\cite{HH} in order to model a Newtonian axially-symmetric galactic system. Nevertheless,  it was soon considered as the paradigm of a two-dimensional (2D)  system that exhibited chaotic behaviour (see, for instance,~\cite{Tabor, Gutzwiller, BoPu}). Later on, when the following generalization containing adjustable parameters was introduced
\begin{equation}
\mathcal{H}=\dfrac{1}{2}(p_{1}^{2}+p_{2}^{2})+ \Omega_{1}  q_{1}^{2}+\Omega_{2} q_{2}^{2}+\alpha \left(
q_{1}^{2}q_{2}+\beta q_{2}^{3}\right) ,
\label{hhmulti}\nonumber
\end{equation}
it was proven that the only Liouville-integrable~\cite{Perelomov} members of this family of generalized   H\'enon--Heiles Hamiltonians
were given by   {\em three} specific choices of the real parameters $\Omega_1$, 
$\Omega_2$, $\alpha$ and $\beta$ (see~\cite{BSV, CTW, GDP, HietarintaRapid, Fordy83, Wojc, SL, FordyHH, Sarlet, RGG, Hindues, Pickering, Conte}  and references therein):

\begin{itemize} 

\item The Sawada--Kotera system, given by $\beta=1/3$ and $\Omega_{1}=\Omega_{2}=\Omega$:
\begin{equation}
 \mathcal{H}=\dfrac{1}{2}(p_{1}^{2}+p_{2}^{2})+ \Omega \left( q_{1}^{2}+ q_{2}^{2}\right)+\alpha \left(
q_{1}^{2}q_{2}+\frac 13 q_{2}^{3}\right) .
\label{HSK1}
 \end{equation}
\item  The Korteweg--de Vries (KdV) system, with $\beta=2$ and $(\Omega_{1},\Omega_{2})$ arbitrary parameters:
\begin{equation}
 \mathcal{H}=\dfrac{1}{2}(p_{1}^{2}+p_{2}^{2})+\Omega_{1}  q_{1}^{2}+\Omega_{2} q_{2}^{2}+\alpha \left(
q_{1}^{2}q_{2}+2 q_{2}^{3}\right).
\label{HKdV1}
 \end{equation}
\item The Kaup--Kuperschdmit system, with $\beta=16/3$ and $\Omega_{2}=16\Omega_{1} =16\Omega$:
\begin{equation}
 \mathcal{H}=\dfrac{1}{2}(p_{1}^{2}+p_{2}^{2})+ \Omega \left( q_{1}^{2}+ 16 q_{2}^{2}\right)+\alpha \left(
q_{1}^{2}q_{2}+\frac {16}3 q_{2}^{3}\right).
\label{HKK1}
 \end{equation}

\end{itemize}
Beyond the former integrable cases, it is worthy to emphasize that there exists a very interesting family of integrable homogeneous potentials, deeply connected to the particular  KdV system (\ref{HKdV1}) that arises when $\Omega_{2}=4\Omega_{1}$. These are the so-called  {Ramani-Dorizzi-Grammaticos (RDG)  series} of integrable potentials~\cite{RDGprl, Hietarinta}, which can be freely superposed  by preserving    integrability,  since they are just the polynomial potentials on the Euclidean plane that can be separated in parabolic coordinates~\cite{Wojc,FF}.  Furthermore, this classical separability property underlies the fact that   a large collection of {\em integrable rational perturbations} can be added to the RDG potentials by preserving the integrability of the complete Hamiltonian (see~\cite{FF,HoneIP,HonePLA,tesis,Annals10} and references therein).

In this paper we  firstly  review the integrable {\em curved} analogue on the 2D sphere \textbf{S}$^{2}$ and the   hyperbolic (or Lobachevski) space  \textbf{H}$^{2}$ of the {\em flat} KdV H\'enon--Heiles Hamiltonian   (\ref{HKdV1}) with $\Omega_{2}=4\Omega_{1}$,  which has been recently presented  in~\cite{HHnon}, together with  the full curved counterpart of the integrable RDG series of potentials. 
In this approach,  all  the results   depend  on the Gaussian curvature $\kappa$ of the underlying space in an explicit form,  so that   all the flat/Euclidean  expressions  can be recovered performing the zero-curvature  limit (contraction) $\k\to 0$ from the curved expressions.  Alternatively, the curvature $\k$ can  also be  understood as a {\em deformation parameter} providing    the  curved systems as   deformed versions  from the flat/Euclidean ones that preserve the integrability of the former. Secondly, we present new integrable perturbations of the curved KdV system, thus generalizing the results of~\cite{HHnon}. 
 
The   structure of the paper  is   as follows.  In the next section, we review all the  {\em flat} integrable Hamiltonian background, that is, the  properties and structure of the KdV H\'enon--Heiles Hamiltonian  $\mathcal{H} $ (hereafter with 
$\Omega_{2}=4\Omega_{1}$) and the RDG series of potentials  on the Euclidean plane  \textbf{E}$^{2}$. 
In section 3,  we construct the known integrable rational perturbations of $\mathcal{H} $ as RDG     potentials with {\em negative} indices.
 In section 4, we  briefly  review  the ambient (or Weierstrass) and Beltrami     (projective) canonical variables for \textbf{S}$^{2}$ and  \textbf{H}$^{2}$, which are  needed  in the curved framework. The resulting {curved} KdV  H\'enon--Heiles Hamiltonian  $\mathcal{H}_\kappa $ and curved RDG potentials  are addressed  in section 5. Finally,  section 6 is devoted to present the new 
 integrable   perturbations of $\mathcal{H}_\kappa $, that can be understood as the  `negative index'  counterpart of the curved RDG potentials, as well as the result that the superposition of all the curved RDG terms does preserve the integrability of the system.

%%%%%%%%%%%%%%%%%%%%%%%%%%%%%%%%%%%%%%%%%%%%%%%

\section{An integrable H\'enon--Heiles system on the plane}

Le us consider the following tuning $\Omega_{2} =4\Omega_{1}=4\Omega  $ in the  KdV H\'enon--Heiles Hamiltonian (\ref{HKdV1}) defined on  \textbf{E}$^{2}$:
\begin{equation}
\mathcal{H} =\frac{1}{2}(p_{1}^{2}+p_{2}^{2})+\Omega (q_{1}^{2}+4q_{2}^{2})+\alpha\left(q_{1}^{2}q_{2}+2 q_{2}^{3}\right) ,\label{KdVflat}
\end{equation}
where $(q_1,q_2)$ are Cartesian coordinates and $(p_1,p_2)$   their conjugate momenta satisfying  the usual canonical Poisson bracket $\{q_{i},p_{j}\}=\delta_{ij}$.
This system  is known to be endowed with a constant of motion  {\em quadratic} in the momenta and given by
\begin{equation}
\mathcal{I} = 
p_{1}(q_{1}p_{2}-q_{2}p_{1})+q_{1}^{2}\left(2\Omega q_{2}+
\dfrac{\alpha}{4}(q_{1}^{2}+4 q_{2}^{2})
\right) ,
 \label{I2KdVflat}
\end{equation}
 that is, $\{ \mathcal{H},\mathcal{I} \}=0$. Hence 
 $\mathcal{H} $ is integrable in the Liouville  sense. Notice that the Hamiltonian  (\ref{KdVflat})   can be regarded as an integrable cubic perturbation added to the $1:2 $ oscillator with frequencies $(\omega, 2\omega)$ once the  identification $\omega^{2}=2\Omega$ is performed.

It is worth making a carefully analysis of the potentials composing both the Hamiltonian (\ref{KdVflat})  and its invariant (\ref{I2KdVflat}). For this, let us recall that the so-called  {\em RDG series  of integrable potentials} consists of homogeneous polynomial potentials of degree $n$   given by~\cite{RDGprl, Hietarinta}
\begin{equation}
\mathcal{V}_{n}(q_1,q_2) =\sum\limits_{i=0}^{[\frac{n}{2}]}2^{n-2i}\dbinom{n-i}{i}q_{1}^{2i}q_{2}^{n-2i}
\, ,\qquad n=1,2,\dots
\label{Ram1}
\end{equation}
In this respect, we remark that  the quadratic potential     (the $1:2 $ oscillator)  and  the 
   cubic potential in $\mathcal{H}$,  say $\mathcal{V}_{2}$ and $\mathcal{V}_{3}$, are just the second- and the third-order  RDG potentials, respectively. Moreover, the integral  $\mathcal{I}$   contains the linear   $ \mathcal{V}_{1}$ and the  quadratic $ \mathcal{V}_{2}$ RDG potentials; namely
 \begin{equation}
 \mathcal{V}_{1}=2 q_{2},\qquad \mathcal{V}_{2}=q_{1}^{2} +4 q_{2}^{2},\qquad  \mathcal{V}_{3}=4q_{1}^2q_{2}+ 8q_{2}^3 .
 \label{Ram}\nonumber
 \end{equation}
 
In general, it can be straightforwardly proven that a Hamiltonian $\mathcal{H}_{n}$ containing the RDG potential  $\mathcal{V}_{n}$  is Liouville integrable, and its integral of the motion $\mathcal{L}_n$  involves  the $\mathcal{V}_{n-1}$ potential, namely
 \begin{equation}
\mathcal{H}_{n}=\dfrac{1}{2}(p_{1}^{2}+p_{2}^{2})+\alpha_{n}\mathcal{V}_{n}, \quad \ 
\mathcal{L}_n=p_{1}(q_{1}p_{2}-q_{2}p_{1})+ \alpha_{n} q_{1}^{2}\mathcal{V}_{n-1} ,\quad\  \{\mathcal{H}_{n},\mathcal{L}_n\}=0 .
\label{Rflat}
\end{equation}
The  formula (\ref{Ram1}) requires the definition of the 0-th order RDG potential   as a trivial constant 
$\mathcal{V}_{0}:=1$, that is, the first Hamiltonian system within the RDG series  reads
 \begin{equation}
\mathcal{H}_{1}=\dfrac{1}{2}(p_{1}^{2}+p_{2}^{2})+\alpha_{1} \left( 2 q_2 \right), \qquad 
\mathcal{L}_1=p_{1}(q_{1}p_{2}-q_{2}p_{1})+ \alpha_{1} q_{1}^{2} .
\label{Rflat1}\nonumber
\end{equation}

A crucial mathematical property of the RDG potentials is the fact that they can be freely superposed without losing   integrability~\cite{Hietarinta,tesis,Annals10}. More explicitly:

\medskip

\noindent
{\bf Proposition 1.}  {\em    The Hamiltonian  written in Cartesian canonical variales $(p_1,p_2,q_1,q_2)$  as 
\begin{equation}
\mathcal{H}_{(M)}=\dfrac{1}{2}\left(
p_{1}^{2}+p_{2}^{2}
\right)+\sum\limits_{n=1}^M
\alpha_n \mathcal{V}_{n} =\dfrac{1}{2}\left(p_{1}^{2}+p_{2}^{2}\right)+\sum\limits_{n=1}^{M}\sum\limits_{i=0}^{[\frac{n}{2}]}\alpha_{n}2^{n-2i}\dbinom{n-i}{i}q_{1}^{2i}q_{2}^{n-2i} ,
\label{bk}
\end{equation}
where $M\in\mathbb{N}^+$ and $\alpha_n$ are arbitrary real constants, 
is endowed with the following  integral of the motion
\begin{eqnarray}
\mathcal{L}_{(M)}\!\!&=&\!\! p_{1}(q_{1}p_{2}-q_{2}p_{1})
+q_{1}^{2}\sum\limits_{n=1}^{M}\alpha_{n}\mathcal{V}_{n-1} \cr
\!\!&=&\!\! p_{1}(q_{1}p_{2}-q_{2}p_{1})
+q_{1}^{2}\left(
\sum\limits_{n=1}^{M}\sum\limits_{i=0}^{[\frac{n-1}{2}]}\alpha_{n}2^{n-1-2i}\dbinom{n-1-i}{i}q_{1}^{2i}q_{2}^{n-1-2i}\right).\label{I2Ramani}
\end{eqnarray}

}

\medskip

Therefore, the relationship between the KdV H\'enon--Heiles Hamiltonian $ \mathcal{H} $ (\ref{KdVflat}) and its constant of motion $ \mathcal{I}$ (\ref{I2KdVflat}) with the Hamiltonian $\mathcal{H}_{(M)} $  (\ref{bk}) and the integral  $\mathcal{L}_{(M)}$ (\ref{I2Ramani})  comes out as a byproduct of proposition 1, since by setting
\begin{equation}
M=3,\qquad \alpha_1=0,\qquad 
\alpha_2=\Omega,\qquad  \alpha_3= \alpha/4 ,\label{const}
\end{equation}
we obtain that
\begin{eqnarray}
&& \mathcal{H} \equiv \mathcal{H}_{(3)} =\dfrac{1}{2}(p_{1}^{2}+p_{2}^{2})+\alpha_{2} \mathcal{V}_{2}+\alpha_{3} \mathcal{V}_{3} ,
\nonumber\\
&&  \mathcal{I} \equiv \mathcal{L}_{(3)}=  
p_{1}(q_{1}p_{2}-q_{2}p_{1})+q_{1}^{2}\left(\alpha_2\mathcal{V}_{1} +
 {\alpha_3} \mathcal{V}_{2}   
\right)  .
\label{be}\nonumber
\end{eqnarray}

%%%%%%%%%%%%%%%%%%%%%%%%%%%%%%%%%%%%%%%%%%%%%%%

\section{Integrable rational perturbations of a  KdV H\'enon--Heiles\\ system on the plane}

It is worth stressing that the RDG potentials can be extended in order to provide integrable rational perturbations of the Hamiltonian (\ref{KdVflat})  by   starting  from the 0-th order RDG potential and going `backwards', that is,  by considering {\em negative} indices $n$.

In order to make this statement explicit, let us start from the trivial Hamiltonian defined by  the 0-th order potential $\mathcal{V}_{0}:=1$. This is clearly integrable as Poisson-commutes with the integral 
$\mathcal{L}_0$ given by
 \begin{equation}
\mathcal{H}_{0}=\dfrac{1}{2}(p_{1}^{2}+p_{2}^{2})+\alpha_{0}  \mathcal{V}_{0}, \qquad 
\mathcal{L}_0=p_{1}(q_{1}p_{2}-q_{2}p_{1})+ \alpha_{0} 1 .
\label{Rflat0} 
\end{equation}
Therefore, according to (\ref{Rflat}), it seems natural to   define 
 \begin{equation}
\mathcal{L}_0:=p_{1}(q_{1}p_{2}-q_{2}p_{1})+ \alpha_{0}q_1^2 \mathcal{V}_{-1} ,\qquad\mbox{where}\qquad  \mathcal{V}_{-1}:=\frac{1}{q_1^2} \equiv   \frac{\mathcal{V}_{0}}{q_1^2} .
\label{Rflat00}\nonumber
\end{equation}
It is quite remarkable  that $ \mathcal{V}_{-1}$ is just a Rosochatius    or  Winternitz potential~\cite{WSUF65,LetN, Non}.  From it, we can construct the corresponding $n=-1$ potential which is again integrable; namely
 \begin{equation}
\mathcal{H}_{-1}=\dfrac{1}{2}(p_{1}^{2}+p_{2}^{2})+\alpha_{-1}  \mathcal{V}_{-1}, \quad\  
\mathcal{L}_{-1}=p_{1}(q_{1}p_{2}-q_{2}p_{1})+ \alpha_{-1}q_1^2 \mathcal{V}_{-2}  ,\quad \  \mathcal{V}_{-2}:=-\frac{2 q_{2}}{q_1^4} \equiv -\frac{ \mathcal{V}_1}{q_1^4} .
\label{Rflatm1}\nonumber
\end{equation}
In this way, the complete series of rational perturbations of the Hamitonian (\ref{KdVflat}), which can be  understood as the  `negative'  counterparts of the RDG potentials  (\ref{Ram1}),   are found    to be~\cite{FF,HonePLA}:
\begin{eqnarray}
&&\mathcal{H}_{-n}=\dfrac{1}{2}(p_{1}^{2}+p_{2}^{2})+\alpha_{-n}\mathcal{V}_{-n}, \qquad 
\mathcal{L}_{-n}=p_{1}(q_{1}p_{2}-q_{2}p_{1})+ \alpha_{-n} q_{1}^{2}\mathcal{V}_{-(n+1)} , \nonumber \\[2pt]
&& 
\mathcal{V}_{-n}=(-1)^{n+1} \dfrac{\mathcal{V}_{n-1}}{q_{1}^{2n}} ,\qquad  \{\mathcal{H}_{-n},\mathcal{L}_{-n}\}=0 , \qquad n=1,2\dots 
\label{Rflatm}
\end{eqnarray}
Nevertheless, for the sake of clarity, let us consider real parameters  with positive indices, $\lambda_n$, defined by
\begin{eqnarray}
\lambda_n:= (-1)^{n+1}\alpha_{-n},\qquad n=1,2\dots
\label{lam}
\end{eqnarray}
which allow us to rewrite (\ref{Rflatm}) as
\begin{eqnarray}
\mathcal{H}_{-n}=\dfrac{1}{2}(p_{1}^{2}+p_{2}^{2})+\lambda_{n}\frac{ \mathcal{V}_{n-1}}{q_1^{2n}}, \quad \ 
\mathcal{L}_{-n}=p_{1}(q_{1}p_{2}-q_{2}p_{1})- \lambda_{n}  \,  \frac{ \mathcal{V}_{n}}{q_1^{2n}} ,\quad\ n=1,2\dots
\label{Rflatm2}
\end{eqnarray}

  Moreover,      the RDG potentials  (\ref{Ram1}), the 0-th potential  $V_0=1$  and the rational perturbations (\ref{Rflatm2}) can be freely superposed leading to an integrable Hamiltonian that generalizes the results  given in proposition~1 as follows.
 \medskip

\noindent
{\bf Proposition 2.}~\cite{Annals10}  {\em    The Hamiltonian  given by 
\begin{eqnarray}
 \mathcal{H}_{(M, R)}\!\!\!&= &\!\! \! \dfrac{1}{2}\left(p_{1}^{2}+p_{2}^{2}\right)+\sum\limits_{n=1}^{M}\alpha_{n}\mathcal{V}_{n}+ {\alpha_0}\mathcal{V}_{0}+\sum\limits_{n=1}^{R}\lambda_{n}\dfrac{\mathcal{V}_{n-1}}{q_{1}^{2n}}
\nonumber\\[2pt]
\!\!\!&= &\!\! \! \dfrac{1}{2}\left(p_{1}^{2}+p_{2}^{2}\right) +\sum\limits_{n=1}^{M}\sum\limits_{i=0}^{[\frac{n}{2}]}\alpha_{n}2^{n-2i}\dbinom{n-i}{i}q_{1}^{2i}q_{2}^{n-2i} + \alpha_0\notag\\
&&\qquad  +
\sum\limits_{n=1}^{R}\sum\limits_{i=0}^{[\frac{n-1}{2}]} \lambda_{n} 2^{n-1-2i}\dbinom{n-1-i}{i} 
\, \dfrac{q_{2}^{n-1-2i}}{q_{1}^{2(n-i)}}  \, ,
\label{KdVFP} 
\end{eqnarray}  
where   $\alpha_n$, $\alpha_0$ and $\lambda_n$ are arbitrary real constants, is integrable for any indices  $M,R\in\mathbb{N}^+$.
The corresponding integral of the motion   reads
\begin{eqnarray}
 \mathcal{L}_{(M, R)}\!\!\!&= &\!\! \!  p_{1}(q_{1}p_{2}-q_{2}p_{1}) +q_{1}^{2} 
\sum\limits_{n=1}^{M}\alpha_{n}\mathcal{V}_{n-1}  + \alpha_0  {\mathcal{V}_{0}}  -\sum\limits_{n=1}^{R}\lambda_{n}\dfrac{\mathcal{V}_{n}}{q_{1}^{2n}}
 \,  ,\label{KdVFPI}\nonumber
\end{eqnarray} 
where $\mathcal{V}_{n}$ are given in (\ref{Ram1}) and $\mathcal{V}_{0}=1$.}

\medskip

For instance, if we set $M=3$ and $R=4$ we obtain the following integrable generalization of the KdV    H\'enon--Heiles Hamiltonian (\ref{KdVflat}):
\begin{eqnarray}
 \mathcal{H}_{(3, 4)}\!\!\!&= &\!\! \!   \dfrac{1}{2}\left(p_{1}^{2}+p_{2}^{2}\right)+\alpha_{1}\mathcal{V}_{1}+ \alpha_{2}\mathcal{V}_{2}+ \alpha_{3}\mathcal{V}_{3}+ \alpha_{0}\mathcal{V}_{0}+  \lambda_{1}\dfrac{\mathcal{V}_{0}}{q_{1}^{2}}+  \lambda_{2}\dfrac{\mathcal{V}_{1}}{q_{1}^{ 4}}+\lambda_{3}\dfrac{\mathcal{V}_{2}}{q_{1}^{ 6}} +  \lambda_{4}\dfrac{\mathcal{V}_{3}}{q_{1}^{ 8}} \nonumber\\[2pt]
 \!\!\!&= &\!\! \!   \dfrac{1}{2}\left(p_{1}^{2}+p_{2}^{2}\right)+\alpha_{1} \left(2 q_{2}  \right)
 + \alpha_{2}  \left( q_{1}^{2} +4 q_{2}^{2}  \right)+ \alpha_{3}  \left(  4q_{1}^2q_{2}+ 8q_{2}^3\right) + \alpha_{0}\nonumber\\[2pt]
&&  \qquad   +  \lambda_{1}\dfrac{ 1}{q_{1}^{2}}+ \lambda_{2}\dfrac{ 2 q_{2}}{q_{1}^{ 4}}+ \lambda_{3}\, \dfrac{ q_{1}^{2} +4 q_{2}^{2}}{q_{1}^{ 6}} +  \lambda_{4}\, \dfrac{ 4q_{1}^2q_{2}+ 8q_{2}^3}{q_{1}^{ 8}} \, , \nonumber
\end{eqnarray}
 which Poisson-commutes with
\begin{eqnarray}
 \mathcal{L}_{(3, 4)}\!\!\!&= &\!\! \!  p_{1}(q_{1}p_{2}-q_{2}p_{1}) +q_{1}^{2}\left(
 \alpha_{1}\mathcal{V}_{0}+ \alpha_{2}\mathcal{V}_{1}+ \alpha_{3}\mathcal{V}_{2}\right) + \alpha_{0}\mathcal{V}_{0}\nonumber\\[2pt]
&&\qquad  -\left( \lambda_{1}\dfrac{\mathcal{V}_{1}}{q_{1}^{2}}+  \lambda_{2}\dfrac{\mathcal{V}_{2}}{q_{1}^{ 4}}+\lambda_{3}\dfrac{\mathcal{V}_{3}}{q_{1}^{ 6}} +  \lambda_{4}\dfrac{\mathcal{V}_{4}}{q_{1}^{ 8}} \right)  \,  . \nonumber
 \end{eqnarray} 
Recall that the $\lambda_1$-potential behaves as a centrifugal barrier on \textbf{E}$^{2}$ when $\lambda_1>0$~\cite{Non}.

%%%%%%%%%%%%%%%%%%%%%%%%%%%%%%%%%%%%%%%%%%%%%%%

\section{Ambient and Beltrami canonical variables} 

In order to achieve   the generalization of the above results to  the 2D sphere  \textbf{S}$^{2}$ and hyperbolic space \textbf{H}$^{2}$, let us consider the   one-parameter family of 3D real Lie algebras   $\mathfrak{so}_{\kappa}(3)$ with commutation relations and Casimir invariant given by~\cite{Non,Non2}:
\begin{equation}
  [J_{12},J_{01}]=J_{02},\qquad [J_{12},J_{02}]=-J_{01},\qquad [J_{01},J_{02}]=\kk J_{12}  , \label{ca}
 \end{equation}
\begin{equation}
  {\cal C}=J_{01}^2+J_{02}^2+\kk J_{12}^2,
 \label{caa}
 \end{equation}
where $\k$ is a real parameter.    The   2D     homogeneous space $  {\rm  SO}_{\k}(3)/{\rm  SO}(2) $,  where
  ${\rm SO}_{\k}(3)$ is the Lie group of   $\mathfrak{so}_{\kappa}(3)$ and ${\rm  SO}(2)=\langle J_{12}\rangle $,  has   constant Gaussian curvature equal to $\k$.   This generic family of homogeneous space comprises the three relevant cases with constant curvature:
   $$
\begin{array}{lll}
\kk>0:\ \mbox{Sphere}&\qquad \kk=0:\ \mbox{Euclidean plane}&\qquad \kk<0:\ \mbox{Hyperbolic space}\\[2pt]
  {\mathbf S}^2={\rm SO}(3)/{\rm  SO}(2)&\qquad  {\mathbf E}^2={\rm  ISO}(2)/{\rm  SO}(2)&\qquad {\mathbf H}^2= {\rm  SO}(2,1)/{\rm SO}(2)
\end{array}
$$
 These 2D spaces can be  embedded in   $\mathbb R^3=(x_0,x_1,x_2)$ where the {\em ambient} or Weierstrass coordinates must satisfy  \begin{equation}
 x_{0}^{2}+\kappa (x_{1}^{2}+x_{2}^{2})=1 .\label{Wcoor}\nonumber
\end{equation}
   Next, if we apply a central projection with pole  
$(0,0,0)\in \mathbb R^{3}$ from  $(x_0,x_1,x_2)\in \mathbb R^3$ to the 2D projective space,  we obtain the
   {\em Beltrami}   coordinates $\bq=(\tq_1,\tq_2)\in \mathbb R^2$  given by
   \begin{equation}
x_0 =\frac{1}{\sqrt{1+ \kk \btq^2}},\qquad
\>x =\frac{\btq}{\sqrt{1+ \k \btq^2}},\qquad \>q=\frac{\>x}{x_0},
\label{Beltr}
\end{equation}
such that  $\>x=(x_1,x_2)$,   the conjugate Beltrami momenta  are $\bp=(p_1,p_2)$ and hereafter  we    denote
$$
\bq^2=q_1^2+q_2^2,\qquad \bp^2=p_1^2+p_2^2,\qquad \bq\cdot \bp=q_1 p_1+ q_2 p_2  .
\label{bba}
$$

A symplectic realization of  $\mathfrak{so}_\kk(3)$ (\ref{ca})  in terms of the Beltrami  canonical variables $(\bq,\bp)$  turns out to be~\cite{LetN,Non,Non2} 
 \begin{equation}
J_{0i}= p_i+\kk (\bq\cdot\bp) q_i,   \quad i=1,2; \qquad J_{12}=q_1 p_2 - q_2 p_1 .
  \label{cd}
 \end{equation}
In this framework, the curved kinetic energy  $ {\cal T}_\k$ for a particle moving on these spaces comes from   the Casimir (\ref{caa}) under the above realization:
 \begin{equation}
 {\cal T}_\k\equiv\frac 12 {\cal C}=\frac 12 (J_{01}^2+J_{02}^2+\kk J_{12}^2)=\frac 12 \left(1+\kk \bq^2\right) \left(\bp^2+\kk(\bq\cdot\bp)^2 \right) .
 \label{ce}
 \end{equation}
Notice that the flat/Euclidean limit $\k\to 0$ of the above expressions  leads to
$$
 x_0=1,\qquad \>x = \btq,\qquad J_{0i}=p_i,\qquad J_{12}= q_1 p_2 - q_2 p_1,\qquad  {\cal T}= \frac 12 \bp^2.
$$

%%%%%%%%%%%%%%%%%%%%%%%%%%%%%%%%%%%%%%%%%%%%%%%

\section{A KdV H\'enon--Heiles system  on the sphere and  the\\ hyperbolic space}

In this section we summarize the construction of   the curved counterpart  of the KdV H\'enon--Heiles system (\ref{KdVflat}), that has been recently presented in~\cite{HHnon} by making use of of ambient and Beltrami dynamical variables. 
Such construction requires  to obtain, firstly, the curved RDG potentials $\mathcal{V}_{\k, n}$ and, secondly, their superposition, so generalizing proposition 1 to the curved case. In this way, the 
definition of the curved integrable KdV H\'enon--Heiles Hamiltonian $\mathcal{H}_\kappa$  comes out as a byproduct, and  the main object in the construction here presented turns out to be the curved RDG potentials.

\medskip

\noindent
{\bf Proposition 3.}~\cite{HHnon} {\em  The   RDG potentials on   the sphere    ${\>S}^{2}$ and   the  hyperbolic space  ${\>H}^{2}$ are defined in Beltrami  coordinates $(q_1,q_2)$ (\ref{Beltr})  as
\begin{equation}
\mathcal{V}_{\k, n}  =\left(\dfrac{1+\kappa   {\>q}^{2}}{1-\k q_{2}^{2}}\right)^{2}
\sum\limits_{i=0}^{[\frac{n}{2}]}2^{n-2i}\dbinom{n-i}{i}  \! \left(
\dfrac{q_{1}}{\sqrt{1+\kappa   {\>q}^{2}}}
\right)^{2i} \!\! \left(
1-\dfrac{i }{n-i}\left[\dfrac{\kappa  q_{1}^{2}}{1+\kappa   {\>q}^{2}}\right]
\right)
\!\left(
\dfrac{q_{2}}{1+\kappa   {\>q}^{2}}
\right)^{n-2i}   
\label{fe}
\end{equation}
 with  $n=1,2\dots$ Each RDG Hamiltonian 
 \begin{equation}
 \mathcal{H}_{\k,n}  = {\cal T}_\k   +\alpha_{n} \mathcal{V}_{\k,n} ,\nonumber
 \end{equation}
is integrable, as  is endowed  with a constant of motion $\mathcal{L}_{\k,n}$ which is quadratic in the momenta 
 \begin{equation}
\mathcal{L}_{\k,n} =J_{01}J_{12}+\alpha_{n} \,\dfrac{q_{1}^{2}}{1+ \kappa \bf{q}^{2} } \mathcal{V}_{\k, n-1} , \qquad \{ \mathcal{H}_{\k,n} ,\mathcal{L}_{\k,n}\}=0 ,
\label{fh}
\end{equation}
where $\mathcal{V}_{\k, 0} $ is defined by
 \begin{equation}
  \mathcal{V}_{\k,0}:=\dfrac{(1+ \kappa q_{2}^{2})(1+ \kappa \>q^2  )}{\left(1- \kappa q_{2}^{2}\right)^{2}} \, ,
\label{fd}
\end{equation}
  and   $J_{01}$,  $J_{12}$  and  ${\cal T}_\k$  are the functions   given by (\ref{cd}) and (\ref{ce}).}
\medskip

We stress that the flat limit $\kappa\rightarrow 0$   of  the above expressions leads to  the Euclidean  RDG potential  $\mathcal{V}_{n}$~(\ref{Ram1}), 
the Hamiltonian  $\mathcal{H}_{n}$ and its  integral  of motion $\mathcal{L}_{n}$ (\ref{Rflat}) along with $\mathcal{V}_{0}=1$. Recall  also that, under the flat limit, Beltrami coordinates reduce to Cartesian ones. However, note that in the curved case $ \mathcal{V}_{\k,0}$ is no longer   a
 trivial potential. 
We also remark that the quadratic RDG Hamiltonian,  $ \mathcal{H}_{\k,2}  = {\cal T}_\k   +\alpha_{2} \mathcal{V}_{\k,2} $,  is just   the    {\em superintegrable}  curved $1:2$ oscillator, formerly   introduced in~\cite{RS} and further studied in~\cite{Non,Non2}.

  In terms of the  ambient coordinates $(x_0,x_1,x_2)$  (\ref{Beltr}), the curved RDG potentials  (\ref{fe}) and (\ref{fd})   turn out to be
  \begin{eqnarray}
\mathcal{V}_{\k, 0}  =  \dfrac{1- \kappa x_{1}^{2}}{(x_{0}^{2}-\kappa x_{2}^{2})^{2}} , \ \   \mathcal{V}_{\k, n}   =
\dfrac{1}{(x_{0}^{2}-\kappa x_{2}^{2})^{2}}\sum\limits_{i=0}^{[\frac{n}{2}]}2^{n-2i}\dbinom{n-i}{i}x_{1}^{2i}\left(
1-\dfrac{i}{n-i}\kappa  x_{1}^{2}
\right) \left(x_{0} x_{2}\right)^{n-2i} 
\label{qq}
\end{eqnarray}
which affords  their parametrization   in any coordinate system under the appropriate change of variables.   

%%%%%%%%%%%%%% figure 1%%%%%%%%%%%%%%%%%%%%%%%%%%

\begin{figure}[ht]
\setlength{\unitlength}{1mm}
\begin{picture}(140,132)(0,0)
\label{figure1}
\footnotesize{
\put(11,70){\includegraphics[scale=0.15]{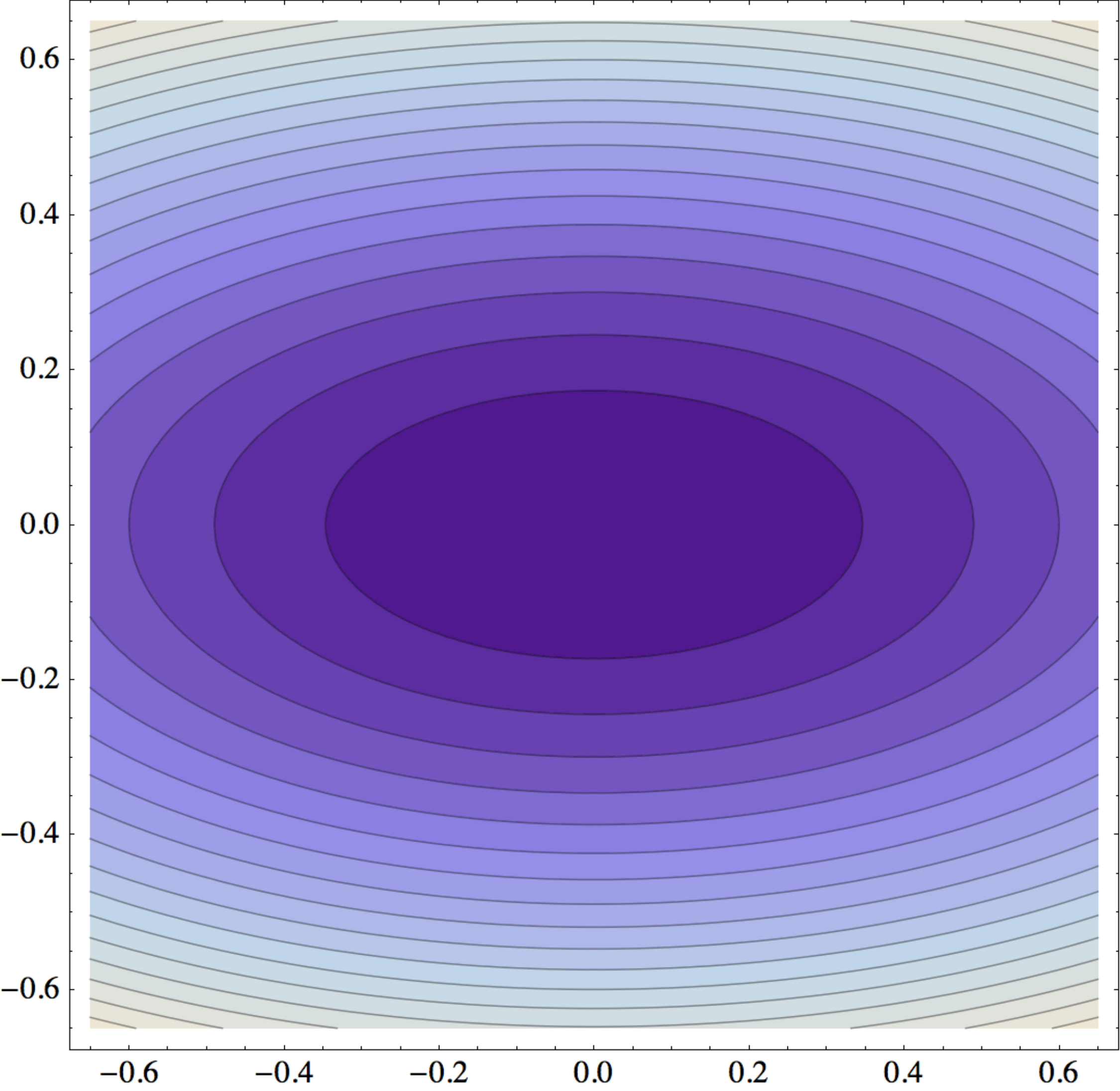}}
\put(38,67){\footnotesize $q_1$}
\put(70,100){\footnotesize $q_2$}
\put(5,122){\footnotesize ${\cal V}_\k$}
\put(5,72){(a)}
\put(85,70){\includegraphics[scale=0.15]{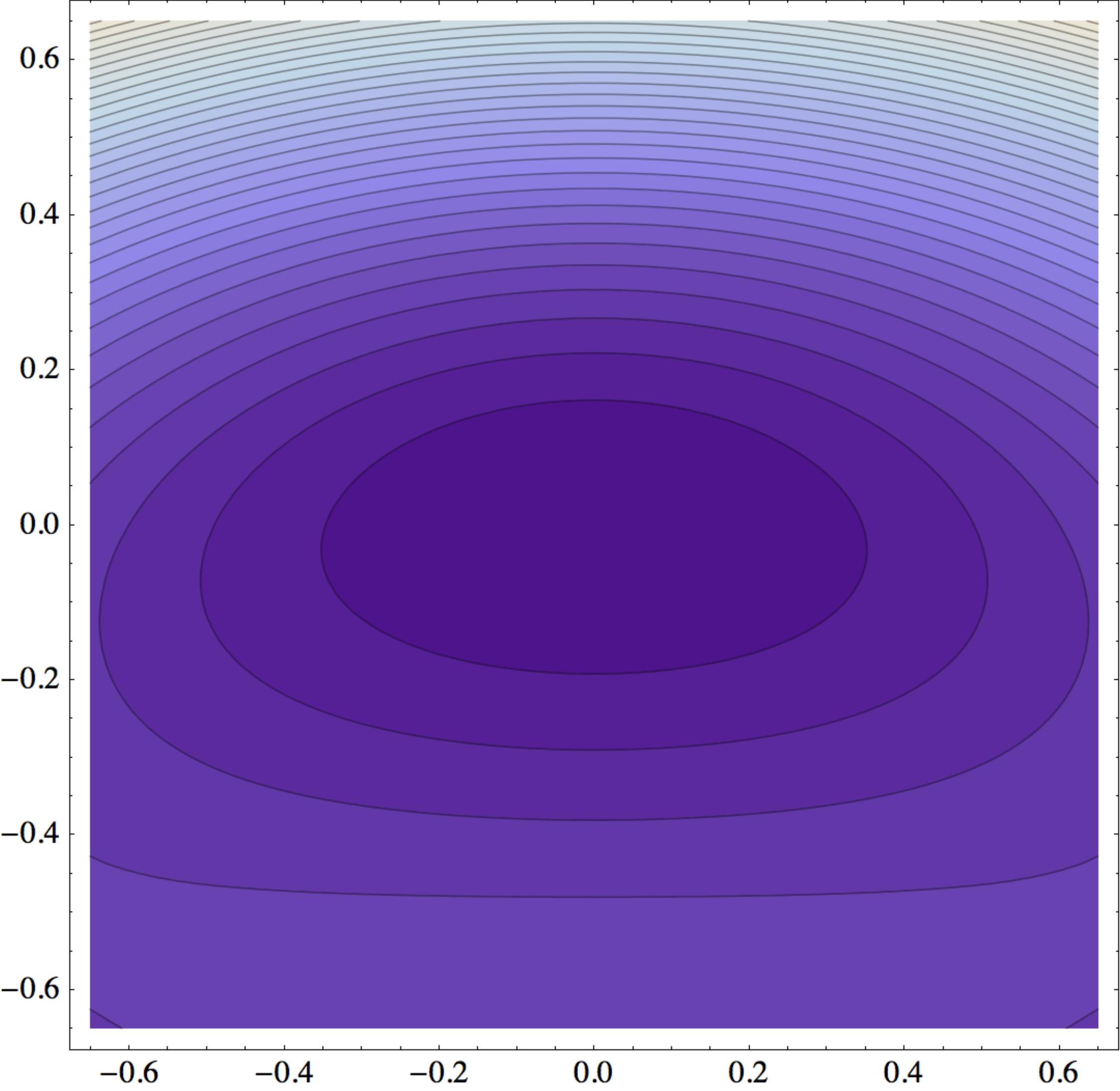}}
\put(115,67){\footnotesize $q_1$}
\put(145,100){\footnotesize $q_2$}
\put(79,122){\footnotesize ${\cal V}_\k$}
\put(79,72){(b)}
\put(11,2){\includegraphics[scale=0.15]{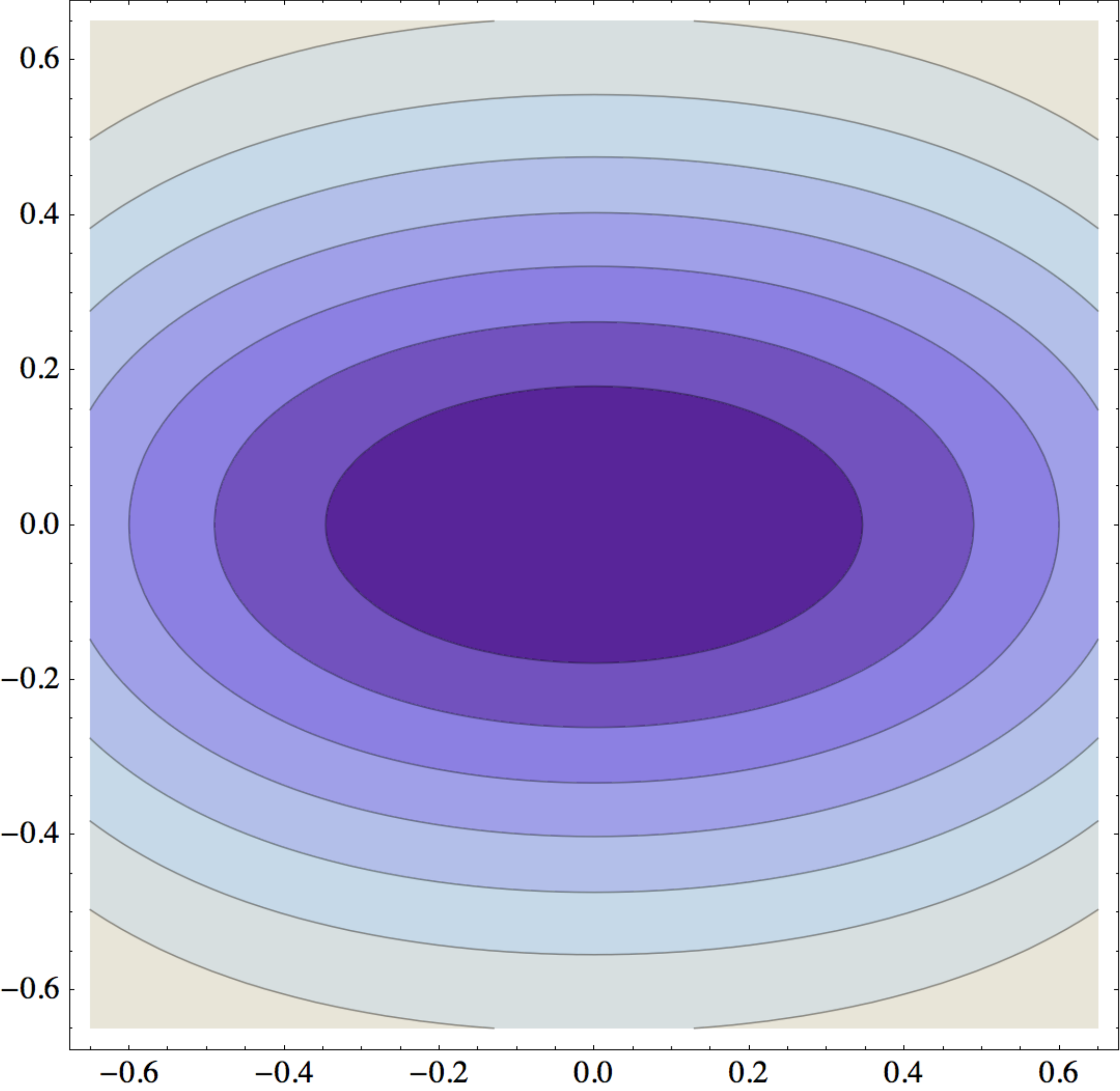}}
\put(38,-2){\footnotesize $q_1$}
\put(69,30){\footnotesize $q_2$}
\put(5,53){\footnotesize ${\cal V}_\k$}
\put(5,3){(c)}
\put(85,2){\includegraphics[scale=0.15]{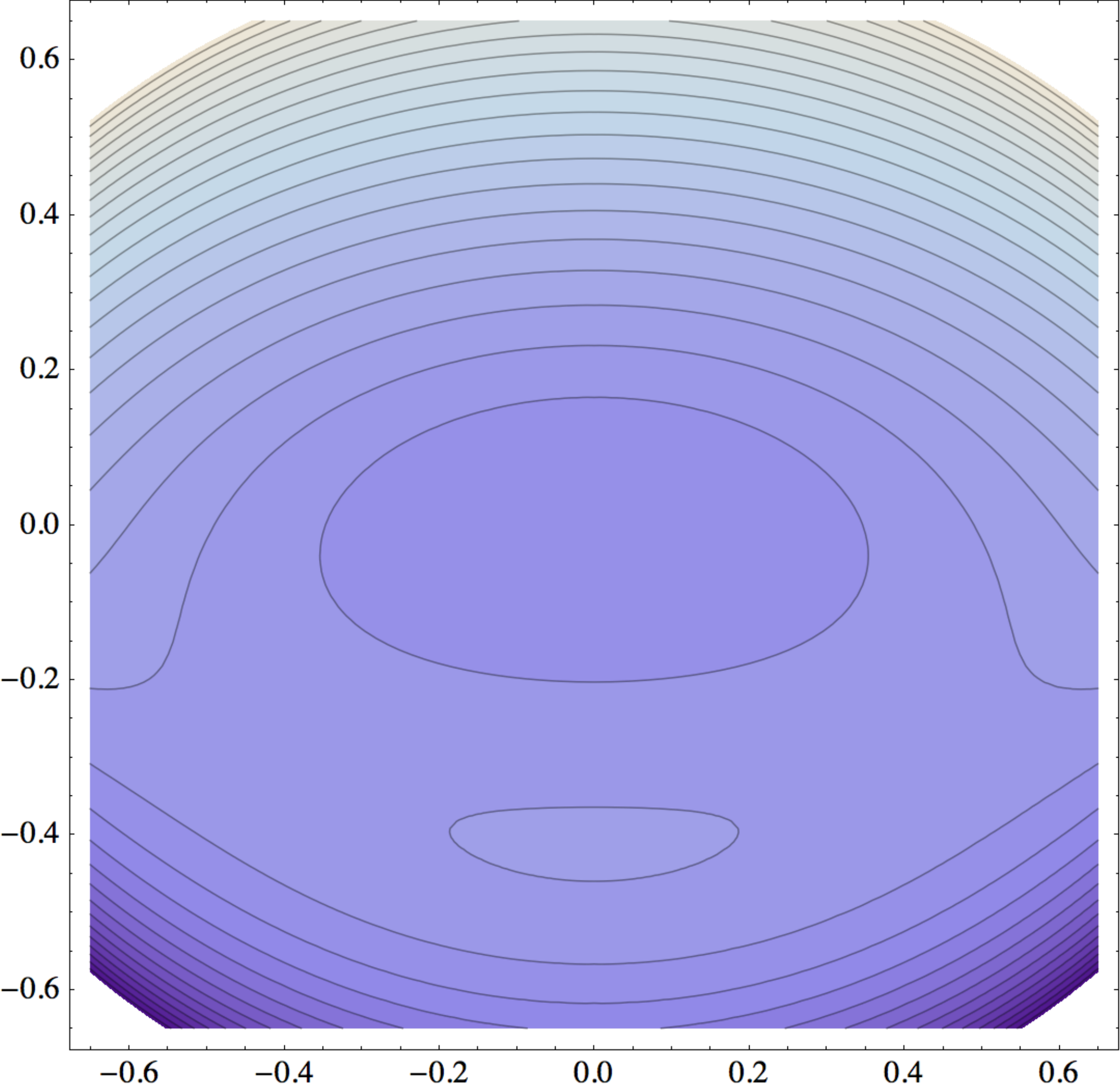}}
\put(115,-2){\footnotesize $q_1$}
\put(144,30){\footnotesize $q_2$}
\put(79,53){\footnotesize ${\cal V}_\k$}
\put(79,3){(d)}
}
\end{picture}
\caption{ \footnotesize
Level plots for the potential ${\cal V}_\k$~\eqref{bex1} in Beltrami projective coordinates $(q_1,q_2)$ for four different values of the constants $(\kappa,\Omega,\alpha)$ involved. Values of the potential function ${\cal V}_\k$ are coded as follows: dark blue represents low values, and higher values are red--shifted. Figure (a) corresponds to $(\kappa,\Omega,\alpha)=(0,1,0)$, (b) to $(0,1,2)$, (c) to $(-1,1,0)$ and (d) to $(-1,1,2)$.
}
\end{figure}
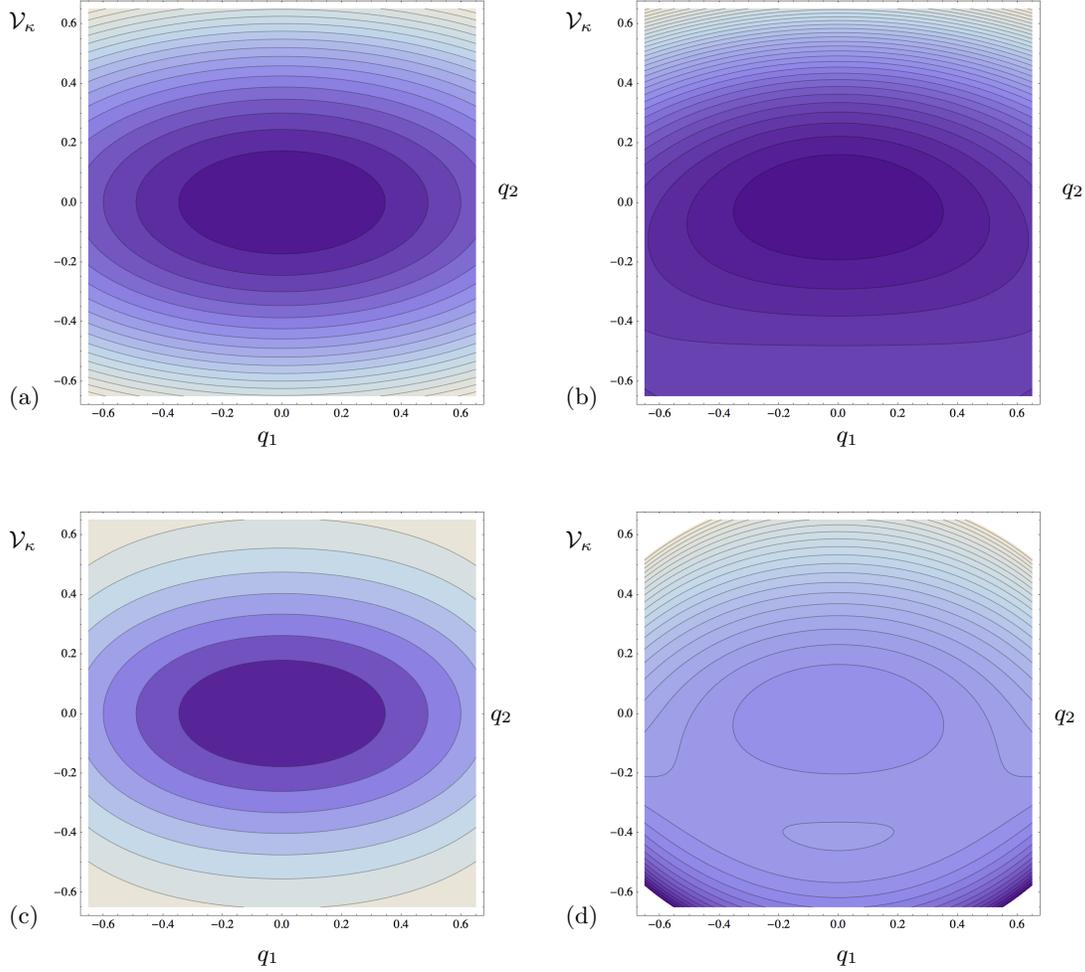

%%%%%%%%%%%%%%%%%%%%%%%%%%%%%%%%%%%%%%%%%%%%%%%%%%%

Furthermore, as in the Euclidean  case, the curved RDG potentials can be freely superposed. Therefore,  the expressions (\ref{bk}) and (\ref{I2Ramani})  given in proposition 1 can be generalized to the curved case as follows.

 \medskip
\noindent
{\bf Proposition 4.}~\cite{HHnon} {\em The Hamiltonian formed by the linear superposition of   the curved RDG potentials (\ref{fe}) and  given by
\begin{equation}
\mathcal{H}_{\kappa,(M)}= {\cal T}_\k +\sum\limits_{n=1}^{M}\alpha_{n}\mathcal{V}_{\k,n} \, , \qquad M\in\mathbb{N}^+ ,
\label{hM}
 \end{equation}
Poisson-commutes with  the function
\begin{equation}
\mathcal{L}_{\kappa,(M)}= J_{01}J_{12}+\dfrac{q_{1}^{2}}{1+ \kappa  \bf{q}^{2}}    \sum\limits_{n=1}^{M}\alpha_{n}\mathcal{V}_{\k, n-1} \, , 
\nonumber
\end{equation}
where $J_{01}$,  $J_{12}$  and  ${\cal T}_\k$  are the functions   given by (\ref{cd}) and (\ref{ce}).
}
  
   \medskip

As a straightforward consequence,  the curved counterpart of the H\'enon--Heiles KdV Hamiltonian~\eqref{KdVflat} on ${\mathbf S}^2$ and ${\mathbf H}^2$ along with its integral (\ref{I2KdVflat}),  written in Beltrami variables,   can be obtained from proposition 4 for the particular case $\mathcal{H}_{\kappa,(3)}$  by setting (\ref{const}); namely
    \begin{eqnarray}
&& \mathcal{H}_\kappa  ={\cal T}_\k + {\cal V}_\k={\cal T}_\k +\Omega\, \mathcal{V}_{\kappa, 2}+\frac {\alpha}4\, \mathcal{V}_{\kappa, 3} \, ,
\nonumber\\[2pt]
&&{\cal V}_\k= \Om\,   \frac{  q_1^2(1+\k q_2^2) + 4 q_2^2}{(1-\kappa  \tq_2^2)^2  }  +
 {\alpha}\,  \frac{   q_1^2 q_2(1+\k\>q^2- \frac 12 \k q_1^2)  + 2q_2^3}{ (1-\kappa  \tq_2^2)^2 (1+\k \>q^2)  }  \, .
\label{bex1}
\end{eqnarray}
And, therefore, the Hamiltonian $\mathcal{H}_\kappa$ Poisson-commutes with  the corresponding integral of the motion that comes from $\mathcal{L}_{\kappa,(3)}$: \begin{eqnarray}
&&  \mathcal{I}_{\kappa}=  J_{01}J_{12}+\frac{q_{1}^{2}}{1+ \kappa  \bf{q}^{2}} \left(\Omega\,  \mathcal{V}_{\k,1} +
\frac {\alpha}4\, \mathcal{V}_{\k,2}   
\right)\nonumber\\[2pt]
&&\quad\ =   \left(   p_1+\kappa (\bq\cdot\bp) q_1  \right)\left( q_1 p_2 - q_2 p_1\right)  +\frac{q_{1}^{2}}{1+ \kappa  \bf{q}^{2}} \left(\Omega\, \frac{2 q_{2}(1+\kappa  {\>q}^{2})}{(1-\kappa  q_{2}^{2})^{2}}+
 {\alpha} \, \frac{  q_1^2(1+\k q_2^2) + 4 q_2^2}{4(1-\kappa  \tq_2^2)^2  }  
\right)\, .
\label{bex2}\nonumber
\end{eqnarray}
  
Level plots for the potential ${\cal V}_\k$~\eqref{bex1} are shown in figure~1, where  the Euclidean $(\kappa=0)$ and   hyperbolic $(\kappa<0)$ KdV H\'enon--Heiles  potentials are represented and compared.  We stress that both the curvature $\kappa$ and the constant $\alpha$ can be considered as deformation parameters that preserve the integrability of the initial Hamiltonian. Case (a) represents the flat $1:2$ (superintegrable) anharmonic  oscillator potential on  \textbf{E}$^{2}$  with no H\'enon--Heiles term, in particular with $\kappa=0$, $\Omega=1$ and $\alpha=0$. In case (b) the H\'enon--Heiles cubic term is added  on  \textbf{E}$^{2}$ through $\alpha=2$  (we keep $\Omega=1$), while the underlying space is still flat ($\kappa=0$). Here we see that the H\'enon--Heiles term breaks the $q_2\rightarrow - q_2$ symmetry of the anharmonic oscillator. The (superintegrable) curved $1:2$ anharmonic oscillator potential on  \textbf{H}$^{2}$ with no H\'enon--Heiles term ($\kappa=-1$, $\Omega=1$ and $\alpha=0$) is represented in case (c). Now we realize that the non-vanishing curvature modifies the values of the potential, although its general shape around the origin is quite similar to the flat case. Finally, when the curved H\'enon--Heiles term is added ($\kappa=-1$, $ \Omega=1$ and $\alpha=2$) the plot (d) is obtained  on  \textbf{H}$^{2}$. Note that although superintegrability is broken when the H\'enon--Heiles term is considered, both (b) and (d) potentials always generate integrable motions on the hyperbolic plane.

%%%%%%%%%%%%%%%%%%%%%%%%%%%%%%%%%%%%%%%%%%%%%%%

\section{Integrable perturbations of a  curved KdV H\'enon--Heiles\\ system }

 In this last section we present, as new results, the curved analogues  on 
 \textbf{S}$^{2}$ and   \textbf{H}$^{2}$
of the   `negative' counterparts of the  RDG potentials (\ref{Rflatm}) and of their integrable superposition (\ref{KdVFP}).

 As in section 3, let us start from the curved Hamiltonian $ \mathcal{H}_{\kappa,0}$ corresponding to the   potential $\mathcal{V}_{\kappa, 0}$ (\ref{fd}), which, in contrast with the flat case $\mathcal{V}_{0}=1$,  is now a non-trivial function:
 \begin{equation}
\mathcal{H}'_{\kappa, 0}= {\cal T}_\k+\alpha_{0}  \mathcal{V}_{\kappa,  0}=  {\cal T}_\k +  \alpha_{0}\, \dfrac{(1+ \kappa q_{2}^{2})(1+ \kappa \>q^2  )}{\left(1- \kappa q_{2}^{2}\right)^{2}}  \,  .
\label{Rflat0k}\nonumber
\end{equation} 
This Hamiltonian is, in fact,  integrable as it Poisson-commutes with the function
 \begin{equation}
\mathcal{L}'_{\kappa, 0} =J_{01}J_{12}+  \alpha_{0}\, \frac{2 \kappa\, q_1^2 q_{2} }{(1-\kappa  q_{2}^{2})^{2}}
 \, .
\label{Rflat0k2}\nonumber
\end{equation}
According to (\ref{fh}),  let us  define 
 \begin{equation}
\mathcal{L}'_{\kappa, 0}  :=  J_{01}J_{12}+\alpha_{0} \, \dfrac{q_{1}^{2}}{ 1+ \kappa \bf{q}^{2} }\, \mathcal{V}'_{\k,  -1}  ,\qquad  \mathcal{V}'_{\kappa, -1}:= \kappa\, \frac{2  q_{2}(1+ \kappa \bf{q}^{2} ) }{(1-\kappa  q_{2}^{2})^{2}} \equiv \kappa  \mathcal{V}_{\kappa,  1}.
\label{Rflat00k}\nonumber
\end{equation}
By induction, it can be easily shown that
$$
\mathcal{V}'_{\k,  -n}=\kappa\, \mathcal{V}_{\k,  n},\qquad n=1,2,\dots
$$
 and they vanish when $\kappa=0$. Consequently, under this procedure we have obtained  curved Beltrami potentials with negative index, but 
 they are just proportional to   (\ref{fe}) and, therefore, we have not
 obtained any new result concerning integrable perturbations.

However, it turns out that the curved analogue of the rational perturbations (\ref{Rflatm}) can be obtained by starting from the  trivial (free geodesic) curved Hamiltonian with a constant potential $\alpha_0$ (as in (\ref{Rflat0})) which is clearly integrable; explicitly
 \begin{equation}
\mathcal{H}_{\kappa, 0}= {\cal T}_\k+\alpha_{0}    ,\qquad \mathcal{L}_{\kappa, 0}  =  J_{01}J_{12}+\alpha_{0},\qquad \{ \mathcal{H}_{\kappa, 0},\mathcal{L}_{\kappa, 0}\}=0.
\label{Rflat0k3}\nonumber
\end{equation} 
If we again  assume that recurrence (\ref{fh}) should hold, we are led to define  
  \begin{equation}
\mathcal{L}_{\kappa, 0}  :=  J_{01}J_{12}+\alpha_{0} \, \dfrac{q_{1}^{2}}{ 1+ \kappa \bf{q}^{2} } \, \mathcal{V}_{\k,  -1}  ,\qquad  \mathcal{V}_{\kappa, -1}:=  \frac{ 1+ \kappa \bf{q}^{2}  }{ q_1^2}  .
\label{Rflat00k2}\nonumber
\end{equation}
Surprisingly enough, we find that    $\mathcal{V}_{\k,  -1}\equiv 1/x_1^2$ is just the curved Rosochatius    or  Winternitz potential, which    corresponds to a noncentral oscillator potential on the sphere with center (in ambient coordinates) located at $O_1=(0,1,0)$ (see~\cite{LetN, Non, RS, lett} for a detailed discussion).

Now, if we construct the Hamiltonian $\mathcal{H}_{\kappa, -1}= {\cal T}_\k+\alpha_{-1}\mathcal{V}_{\k,  -1}  $,   the potential $\mathcal{V}_{\k,  -2}$ is then obtained through the corresponding integral of motion:
 \begin{equation}
\  \mathcal{L}_{\kappa, -1}   =  J_{01}J_{12}+\alpha_{-1} \, \dfrac{q_{1}^{2}}{ 1+ \kappa \bf{q}^{2} } \, \mathcal{V}_{\k,  -2}  ,\quad  \mbox{where} \quad \mathcal{V}_{\kappa, -2}:= - \frac{ 2 q_2 (1+ \kappa \bf{q}^{2}  ) }{ q_1^4}=- \frac{(1-\kappa q_2^2)^2}{q_1^4}\, \mathcal{V}_{\kappa, 1}  \, .
\label{Rflat0k21}\nonumber
\end{equation} 
From this, the   integrable curved RDG potentials $\mathcal{V}_{\k,  -n}$  can be defined as follows.

\medskip

\noindent
{\bf Proposition 5.}   {\em Let us define
\begin{equation}
\qquad \mathcal{V}_{\kappa, -1}=  \frac{ 1+ \kappa \bf{q}^{2}  }{ q_1^2}  ,\qquad
\mathcal{V}_{\kappa,-m}=(-1)^{m+1} \frac{(1-\kappa q_2^2)^2}{q_1^{2m}}\, (1+ \kappa  { \bf{q}}^{2}   )^{m-2}\,\mathcal{V}_{\kappa, m-1} \, ,
 \label{HSgen3}
\end{equation}
with $m=2,3,\dots$ and $ \mathcal{V}_{\kappa, m-1}$ given by (\ref{fe}).
The Hamiltonian  defined by $\mathcal{H}_{\kappa,-n} =\mathcal{T}_{\kappa}+\alpha_{-n}\mathcal{V}_{\kappa,-n}$  ($n=1,2,\dots$)  Poisson-commutes with
\begin{eqnarray}
\mathcal{L}_{\kappa, -n}=J_{01}J_{12}+\alpha_{-n} \, \dfrac{q_{1}^{2}}{1+ \kappa  \bf{q}^{2}} \,  \mathcal{V}_{\kappa,-(n+1)} \, .
\label{Rflatm3}\nonumber
\end{eqnarray}
}

 Consequently, the generalization of  the Euclidean rational perturbations (\ref{Rflatm})   to a constant curvature framework is achieved. 
In ambient coordinates (\ref{Beltr}), the potentials (\ref{HSgen3}) read
\begin{equation}
   \mathcal{V}_{\k, -1}   = \frac{1}{x_1^2},\qquad    \mathcal{V}_{\k, -m}   =(-1)^{m+1}\, \frac{(x_0^2-\k x_2^2)^2}{x_1^{2m}}\, \mathcal{V}_{\kappa, m-1} ,\qquad m=2,3,\dots
   \label{popo}
\end{equation}
with $\mathcal{V}_{\kappa, m-1}$ given in (\ref{qq}). These results are illustrated in table~1 by writing the first RDG potentials with positive and negative indices on ${\mathbf E}^2$, ${\mathbf S}^2$  and ${\mathbf H}^2$. We also remark that proposition 5 can be written in terms of coefficients with positive indices, $\lambda_n$, through the   definition  (\ref{lam}).

Finally, the potentials  $\mathcal{V}_{\k, -n} $ can also be added to the Hamiltonian (\ref{hM}) leading to the following full integrable curved superposition, that constitutes the main result of this paper. We write it  in ambient coordinates as follows.

 %%%%%%%%%%%%%%%%%%%%%%%%%%%%%%%%%%%%%%%%%%%%%%%%%%%%%%%%%%%%%%%%%%

 \begin{table}[t]

 {\footnotesize{
\caption{{The   RDG potentials    for $n=\{ 0,\pm 1,\pm 2, \pm 3\}$ on ${\mathbf E}^2$ in Cartesian coordinates $\>q$ (\ref{Ram1}) and (\ref{Rflatm}) along with   their curved counterpart on  ${\mathbf S}^2$ and  ${\mathbf H}^2$ in   ambient coordinates (\ref{qq}) and (\ref{popo}) such that  $x_{0}^{2}+\kappa \>x^2=1$. Recall that  $x_0=1$ and  $\>x=\>q$ when $\k=0$. }} 
\vskip0.5em
\label{Table1}
 \begin{center}
\noindent
\begin{tabular}{llll}
\hline
\\[-0.2cm]
\multicolumn{1}{l}{ ${\mathbf E}^2$: Cartesian coordinates $\>q$}&
\multicolumn{1}{l}{\qquad${\mathbf S}^2$ and  ${\mathbf H}^2$:  Ambient coordinates $(x_0,\>x)$}
 \\[0.2cm]
\hline
 \\[-0.2cm]
 $ \mathcal{V}_{-3}=\dfrac {q_{1}^2+ 4q_{2}^2}{q_1^6} $&\qquad$ \displaystyle{ \mathcal{V}_{\k,-3} =\dfrac{x_1^2 (1-\k x_1^2)+  4 x_0^2x_2^2 } { x_1^6}}  $\\[10pt]
  $ \mathcal{V}_{-2}=-\dfrac {2q_2}{q_1^4} $&\qquad$ \displaystyle{ \mathcal{V}_{\k,-2} =- \dfrac{2 x_0x_2 } { x_1^4}}  $\\[10pt]
 $ \mathcal{V}_{-1}=\dfrac 1{q_1^2} $&\qquad$ \displaystyle{ \mathcal{V}_{\k,-1} =\dfrac{1} { x_1^2}}  $\\[8pt]
$ \mathcal{V}_{0}=1 $&\qquad$ \displaystyle{ \mathcal{V}_{\k,0} =\dfrac{1- \kappa x_{1}^{2}}{(x_{0}^{2}-\kappa x_{2}^{2})^{2}}}  $\\[10pt]
$ \mathcal{V}_{1}=2 q_2 $&\qquad$ \displaystyle{  
 {V}_{\k,1} =\dfrac{2 x_0x_2  }{(x_{0}^{2}-\kappa x_{2}^{2})^{2}}}  $\\[10pt]
$ \mathcal{V}_{2}=q_{1}^2+ 4q_{2}^2   $&\qquad$ \displaystyle{  
 {V}_{\k,2}=   \frac{x_1^2 (1-\k x_1^2)+  4 x_0^2x_2^2  }{(x_{0}^{2}-\kappa x_{2}^{2})^{2}}}  $\\[10pt]
$ \mathcal{V}_{3}=4 q_{1}^2 q_2+ 8q_{2}^3   $&\qquad$ \displaystyle{  
 {V}_{\k,3}=   \dfrac{ 4 x_0 x_1^2  x_2 (1-\frac 12 \k x_1^2)+ 8  x_0^3x_2^3  }{(x_{0}^{2}-\kappa x_{2}^{2})^{2}}}  $\\[10pt]
    \hline
\end{tabular}
\end{center}
}}
 \end{table}

%%%%%%%%%%%%%%%%%%%%%%%%%%%%%%%%%%%%%%%%%%%%%%%%%%%%%%%%%%%%%%%%%%

\medskip

\noindent
{\bf Theorem 6.}   {\em The Hamiltonian formed by the linear superposition of   the curved RDG potentials (\ref{qq}) and (\ref{popo}) 
\begin{equation}
\mathcal{H}_{\kappa, (M,\,R)}=\mathcal{T}_{\kappa}+\sum\limits_{n=1}^{M}\alpha_{n}\mathcal{V}_{\kappa,n}+\alpha_{0} \mathcal{V}_{\kappa,0}+ 
\lambda_{1} \, \dfrac{1}{x_{1}^{2}}+
(x_{0}^{2}-\kappa x_{2}^{2})^{2}\sum\limits_{m=2}^{R}\lambda_{m}\dfrac{\mathcal{V}_{\kappa,m-1}}{x_{1}^{2m}}\, ,
\label{HSgen}\nonumber
\end{equation}
where   $\alpha_n$, $\alpha_0$, $\lambda_1$, $\lambda_m$ are arbitrary real constants and $M,R\in\mathbb{N}^+$, is integrable. Its constant of the motion is given by the function
\begin{equation}
\begin{array}{l}
\mathcal{L}_{\kappa,(M,R)}=J_{01} J_{12}+x_{1}^{2}\sum\limits_{n=1}^{M}\alpha_{n}\mathcal{V}_{\kappa,n-1}+\alpha_{0}x_1^2\kappa\mathcal{V}_{\kappa,1}- \lambda_{1} \,\dfrac{2x_{2}x_{0}}{x_{1}^{2}}-(x_{0}^{2}-\kappa x_{2}^{2})^{2}\sum\limits_{m=2}^{R}\lambda_{m}\dfrac{\mathcal{V}_{\kappa,m}}{x_{1}^{2m}}\, .
 \label{ISgen}\nonumber
\end{array}
\end{equation}
}

In this way, we have obtained the generalization of proposition 2 to   ${\mathbf S}^2$ and  ${\mathbf H}^2$. The above result can be straightforwardly written in terms of Beltrami variables through (\ref{Beltr}).  
 
Finally, we would like to mention that the construction of the curved counterpart of the KdV H\'enon--Heiles Hamiltonian (\ref{HKdV1}) with arbitrary   parameters  $(\Omega_{1},\Omega_{2})$, as well as of the Sawada--Kotera  (\ref{HSK1}) and the
 Kaup--Kuperschdmit (\ref{HKK1}) H\'enon--Heiles systems   is currently in  progress.

%%%%%%%%%%%%%%%%%%%%%%%%%%%%%%%%%%%%%%%%%%%%%%%%%%%

\section*{Acknowledgements}

This work was partially supported by the Spanish MINECO under grant MTM2013-43820-P and   by Junta de Castilla y Le\'on  under   grant BU278U14.

%%%%%%%

\end{document}